\documentclass[12pt]{iopart}
\usepackage{iopams}  
\usepackage{graphicx}
\expandafter\let\csname equation*\endcsname\relax
\expandafter\let\csname endequation*\endcsname\relax
\usepackage{amsmath}
\usepackage{xspace}

\newcommand{\CoBa}{BaFe$_{1.85}$Co$_{0.15}$As$_2$\xspace}
\newcommand{\Ba}{BaFe$_2$As$_2$\xspace}
\newcommand{\Eu}{EuFe$_2$As$_2$\xspace}

\begin{document}

\title[Electron-phonon coupling in 122 Fe pnictides by time-resolved photoemission]{Electron-phonon coupling in 122 Fe pnictides analyzed by femtosecond time-resolved photoemission}


\author{L. Rettig$^{1,2}$, R. Cort\'es$^{1,3}$, H.S. Jeevan$^4$, P. Gegenwart$^4$, T. Wolf$^5$,	J. Fink$^6$ and U. Bovensiepen$^2$}

\address{$^1$Fachbereich Physik, Freie Universit\"at Berlin, Arnimallee 14, D-14195 Berlin, Germany}
\address{$^2$Fakult\"{a}t f\"{u}r Physik, Universit{\"a}t Duisburg-Essen, Lotharstr. 1, D-47048 Duisburg, Germany}
\address{$^3$Abt. Physikalische Chemie, Fritz-Haber-Institut d. MPG, Faradayweg 4-6, D-14195 Berlin, Germany}
\address{$^4$I. Physik. Institut, Georg-August Universit\"at G\"ottingen, D-37077 G\"ottingen, Germany}
\address{$^5$Karlsruhe Institute of Technology, Institut f\"ur Festk\"orperphysik, D-76021 Karlsruhe, Germany}
\address{$^6$Leibniz-Institute for Solid State and Materials Research Dresden, P.O.Box 270116, D-01171 Dresden, Germany }
\ead{uwe.bovensiepen@uni-due.de}

\date{Received: date / Accepted: date}

\begin{abstract}
Based on results from femtosecond time-re\-sol\-ved photoemission, we compare three different methods for determination of the electron-phonon coupling constant $\lambda$ in Eu and Ba-based 122 FeAs compounds. We find good agreement between all three methods, which reveal a small $\lambda<0.2$. This makes simple electron-phonon mediated superconductivity unlikely in these compounds.
\end{abstract}

\maketitle

\section{Introduction}
\label{sec:intro}
Albeit an enormous amount of research has been conducted in the past few years on the Fe pnictide high-temperature superconductors (HTSCs)~\cite{Johnston2010,Stewart2011}, the search for the superconducting pairing mechanism in these materials is still ongoing. Besides other excitations such as spin fluctuations~\cite{Mazin2008}, the electron-phonon (e-ph) coupling responsible for Cooper pairing in conventional superconductors in the Bardeen-Cooper-Schrieffer (BCS) theory is considered a potential candidate. Therefore, the quantitative determination of the e-ph coupling constant $\lambda$ in the Fe pnictides is of particular interest. Several methods to determine $\lambda$ have been established in the literature and successfully used in various studies. In the energy domain, angle-resolved photoemission spectroscopy (ARPES) analyzes the renormalization of the single-particle spectral function by the electronic self energy $\Sigma$, which includes the e-ph coupling. However, such type of analysis requires very high sample and data quality, which is difficult to obtain in the Fe pnictides, and only few ARPES studies of FeAs compounds were able to determine effects of $\Sigma$~\cite{Wray2008,Richard2009,Koitzsch2009}. In addition, in thermal equilibrium, contributions from other degrees of freedom to $\Sigma$ are often difficult to disentangle from e-ph coupling.

In contrast, femtosecond (fs) time-resolved spectroscopies allow to separate e-ph scattering from other relaxation channels like e.g. electron-electron (e-e) scattering or heat diffusion due to their different intrinsic timescales out of thermal equilibrium~\cite{Allen1987,DelFatti2000,Lisowski2004}. Here, time-resolved ARPES (trARPES) is used, which combines the energy and momentum resolution of ARPES with fs time resolution into a powerful tool to directly investigate the dynamics of the electronic structure in a non-equilibrium state~\cite{Bovensiepen2010,Bovensiepen2012}.

The e-ph relaxation in metals after photoexcitation has been successfully described using the two-temperature model (2TM)~\cite{Allen1987,DelFatti2000,Lisowski2004,Anisimov1974,Rethfeld2002,Perfetti2007}. This model describes a system of two coupled heat baths for the conduction electrons and the ion lattice with temperatures $T_e$ and $T_l$ by coupled rate equations for $T_e$ and $T_l$, respectively. One key assumption of this model is that e-e and phonon-phonon (ph-ph) scattering occurs on a much faster timescale than the e-ph scattering, leading to a thermal distribution within each subsystem. However, in correlated electron systems like the cuprate or Fe pnictide HTSCs, e-ph scattering might occur on similar timescales~\cite{Gadermaier2010}, which poses some questions on the applicability of the 2TM for these materials and care has to be taken. Therefore, we analyze our data in terms of a suitable version of the 2TM and compare the results with two complementary methods for the determination of the e-ph coupling strength. We find that all three methods to analyze the e-ph coupling strength from time-resolved photoemission data yield results for the second moment of the Eliashberg coupling function $\lambda\left\langle\omega^2\right\rangle$ which agree well within error bars and reproduce the trends for three different 122 FeAs compounds consistently.

\section{Experiments and results}
\label{sec:exp}
For trARPES experiments, single crystals of \Eu and \Ba parent compounds and optimally doped \CoBa ($T_c=23\mathrm{K}$) were cleaved in ultrahigh
vacuum ($p<10^{-10}\,\mathrm{mbar}$) at $T=100\,\mathrm{K}$ where most measurements were carried out. The experimental setup is sketched in fig.~\ref{fig:1}(a): The output of a commercial regenerative Ti:Sapphire amplifier (Coherent RegA 9050) delivering ultrashort laser pulses at $h\nu_1 = 1.5\,\mathrm{eV}$ photon energy with a pulse duration of $55\,\mathrm{fs}$, and operating at $300\,\mathrm{kHz}$ is used to optically excite the samples (pump pulse). A time-delayed frequency-quadrupled probe pulse at $h\nu_2 = 6.0\,\mathrm{eV}$ photon energy with a pulse duration of $80\,\mathrm{fs}$ leads to the emission of photoelectrons, which are detected using an electron time-of-flight (TOF) spectrometer with an acceptance angle of $\pm3^\circ$. The energy resolution of $50\,\mathrm{meV}$ is mainly determined by the spectral width of the probe pulses, and the overall temporal resolution is $<100\,\mathrm{fs}$. For details see~\cite{Lisowski2004}.

\subsection{Three-Temperature model}

\begin{figure}[tb]
\begin{center}
\resizebox{0.9\columnwidth}{!}{\includegraphics{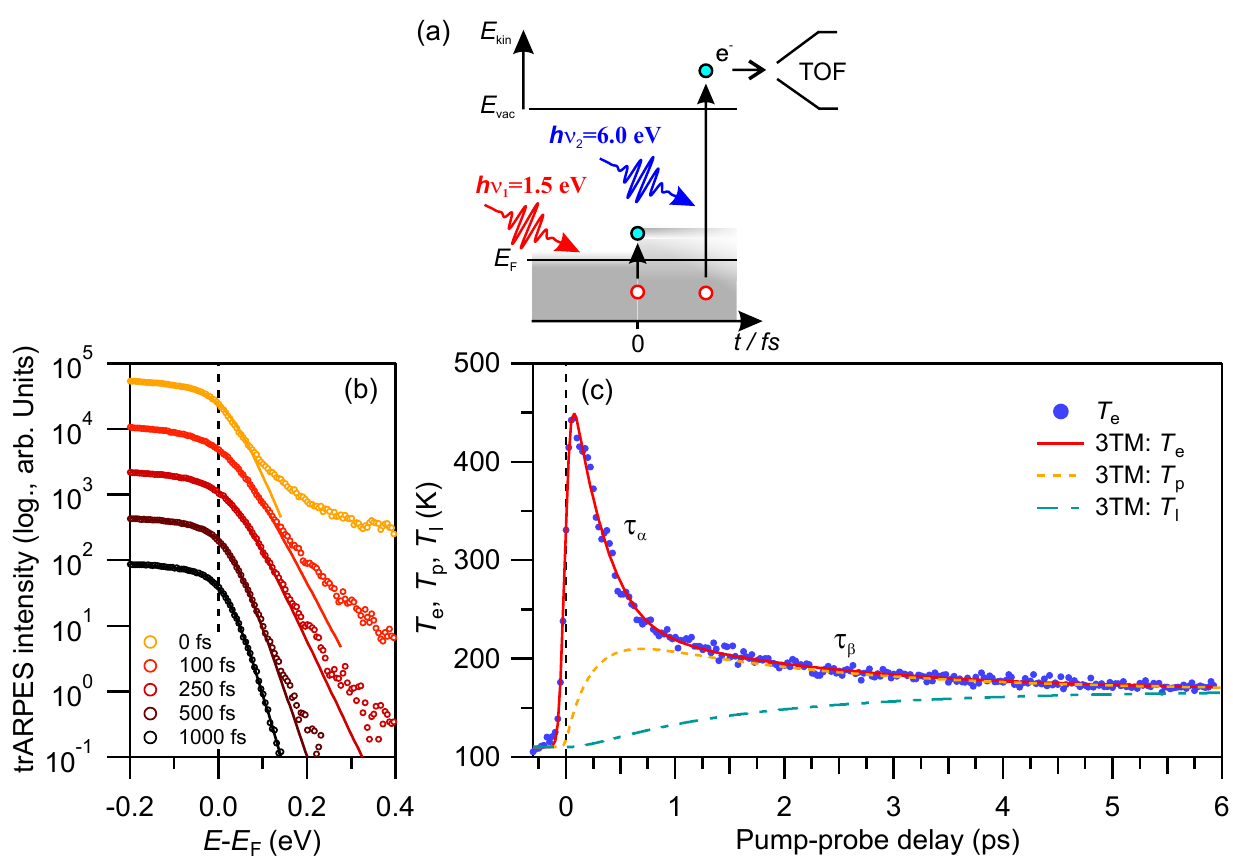}}
\end{center} 
\caption{(a) Scheme of the pump-probe experiment. (b) Time-resolved ARPES spectra of \Eu at $T=100\,\mathrm{K}$ and at normal emission for various pump-probe delays on a logarithmic intensity scale using an incident pumping fluence of $F=0.8\,\mathrm{mJ/cm^2}$. Lines are fits to Fermi-Dirac distribution functions. (c) Electronic temperature $T_e$ and a fit to the 3TM. Temperatures of the hot phonons ($T_p$) and the rest of the lattice modes ($T_l$) are shown as dashed and dash-dotted lines, respectively.}
\label{fig:1}     
\end{figure}

Exemplary trARPES data of \Eu are shown in fig.~\ref{fig:1}(b) for various pump-probe delays on a logarithmic intensity scale. After excitation, a pronounced distribution of excited charge carriers is formed at $E-E_\mathrm{F}>0.1\,\mathrm{eV}$, which deviates from the Fermi-Dirac distribution of the electronic system before excitation and which originates from hot, non-thermalized electrons. Subsequently, e-e scattering quickly leads to thermalization of these non-thermal electrons and to the formation of a hot thermalized electron distribution~\cite{Lisowski2004}. At $t>{250}\,\mathrm{fs}$, the non-thermal contribution has decayed to less than 1\% of the electron population. 

The temperature of the thermalized part of the electronic system $T_e$ can be extracted from the trARPES data by fitting a Fermi-Dirac distribution function to the high-energy cutoff of the transient spectra, multiplied by a phenomenological density-of-states function and convoluted with an instrumental resolution function~\cite{Lisowski2004,Perfetti2007}, as shown in fig.~\ref{fig:1}(b) (solid lines). $T_e$ determined by the fitting is shown in fig.~\ref{fig:1}(c) as a function of pump-probe delay. After the steep rise of the electronic temperature at zero pump-probe delay, we find a relaxation on two distinct timescales $\tau_\alpha$ and $\tau_\beta$. Such a behavior indicates the selective coupling of hot electrons to a subset of strongly coupled phonon modes on the faster timescale $\tau_\alpha$ and the subsequent energy transfer to the rest of the phonons by anharmonic ph-ph scattering, determined by $\tau_\beta$. Thus, the system is described by an extended version of the 2TM, which had been developed for the cuprate HTSCs~\cite{Perfetti2007}, where a similar relaxation dynamics was found, and which has been also used recently for the Fe pnictides~\cite{Mansart2010,Avigo2013}. This three-temperature model (3TM) consists of a system of three coupled differential equations describing the temperature of the electrons, $T_e$, of a hot phonon distribution $T_p$  and of the rest of the lattice modes, $T_l$:
\begin{eqnarray}
\frac{\mathrm{d} T_e}{\mathrm{d} t} &=& -H(T_e, T_p) + \frac{S}{C_e}\\
\frac{\mathrm{d} T_p}{\mathrm{d} t} &=& +\frac{C_e}{C_p}H(T_e, T_p) - \frac{T_p-T_l}{\tau_\beta}\\
\frac{\mathrm{d} T_l}{\mathrm{d} t} &=& +\frac{C_p}{C_l}\frac{T_p-T_l}{\tau_\beta}\qquad.
\end{eqnarray}
The source term $S$ describes the optical excitation, $C_e$, $C_p$ and $C_l$ are the specific heat capacities of electrons, strongly and weakly coupled phonon modes, respectively. The anharmonic decay of phonons is described by $\tau_\beta$. For the energy transfer from the electrons to the more strongly coupled phonons, the formula derived by Allen~\cite{Allen1987} is used:
\begin{equation}
H(T_e, T_p)=\gamma_T (T_e-T_p)=\frac{3\hbar\lambda\left\langle\omega^2\right\rangle}{\pi k_B}\frac{T_e-T_p}{T_e}\qquad,
\end{equation}
where $k_B$ is the Boltzmann constant. This relation allows determination of the second moment of the Eliashberg e-ph coupling function  $\lambda\left\langle\omega^2\right\rangle$. 

A fit of a numerical solution of the 3TM to $T_e$ is shown in fig.~\ref{fig:1}(c) and yields a good agreement to the experimental data. Due to an ambiguity of the model parameters such as the electron and lattice specific heat capacities reported in the literature, a range of values for $\lambda\left\langle\omega^2\right\rangle$ is retrieved, where the fits show similar good agreement to the data. Details of this analysis can be found in~\cite{Avigo2013}; here, only the main results are discussed. From the 3TM, we find values of $\lambda\left\langle\omega^2\right\rangle=56-65\,\mathrm{meV^2}$ for \Eu, while in \CoBa it is slightly smaller ($\lambda\left\langle\omega^2\right\rangle=46-55\,\mathrm{meV^2}$) and even smaller in undoped \Ba ($\lambda\left\langle\omega^2\right\rangle=30-46\,\mathrm{meV^2}$).

\subsection{Electronic excess energy}

\begin{figure}[tb]
\begin{center}
  \resizebox{0.8\columnwidth}{!}{\includegraphics{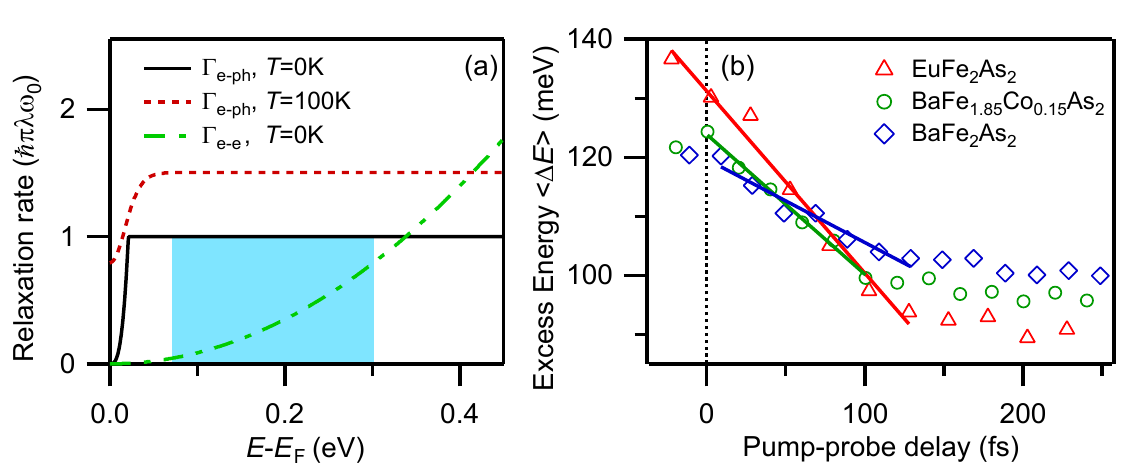}}
\end{center}

\caption{(a) e-ph and e-e contributions to the electron decay rate $\Gamma$. The e-ph contribution $\Gamma_{e-ph}$ calculated in the Debye model with $\hbar\omega_D=20\,\mathrm{meV}$ and $\lambda=0.3$ at $T=0\,\textrm{K}$ and $T=100\,\textrm{K}$ increases up to $\sim\omega_D$ and is constant above. The electronic contribution $\Gamma_{e-e}$ calculated for $\beta=0.1\,\mathrm{eV^{-1}}$ exceeds $\Gamma_{e-ph}$ only at higher energies. The shaded area marks the energy where e-ph scattering dominates. (b) Electron mean excess energy extracted from trARPES data at $T=100\,\mathrm{K}$ near $k_\mathrm{F}$ within the energy window $0.07\,\mathrm{eV}<\epsilon<0.3\,\mathrm{eV}$ marked in (a). Solid lines are fits to equation~\eqref{eqn:e-ph}.}
\label{fig:2}      
\end{figure}

The analysis within the 3TM was based on the energy relaxation within the thermalized part of the transient electronic distribution function and thus neglected the non-thermal electrons present at early delay times (see fig.~\ref{fig:1}(b)). However, the analysis of the energy relaxation of these non-thermal, excited electrons can also provide information about the strength of e-ph coupling.

The scattering rate of electrons excited at energy $\epsilon=E-E_\mathrm{F}$ above the Fermi level is in the self-energy formulation of many-body theory determined by $\Gamma = \hbar\tau^{-1} = 2 \mathrm{Im}\Sigma(\epsilon)$, where $\mathrm{Im}\Sigma$ is the imaginary part of the electronic self energy~\cite{Echenique2000}. Important energy-dependent contributions to $\Gamma$ arise from e-e and e-ph scattering, where e-e scattering is considered to follow the quadratic energy scaling of the Fermi liquid theory, $\Gamma_{e-e}=2\beta\left[(\pi k_B T)^2+ \epsilon^2\right]$ (fig.~\ref{fig:2}(a)). As the proportionality coefficient $\beta$ is rather small in typical metals (in the order of ${10^{-2}}\mathrm{eV^{-1}}$ to ${10^{-1}}\mathrm{eV^{-1}}$)~\cite{McDougall1995,Valla1999}, this contribution is negligible compared to the e-ph scattering at low enough excitation energies. The latter increases up to the maximal phonon energy $\hbar\omega_{max}$ and is constant above (fig.~\ref{fig:2}(a)). For $\epsilon>\hbar\omega_{max}$ and $T=0\,\textrm{K}$, $\Gamma_{e-ph}$ results for an Einstein mode $\omega_0$ to~\cite{Engelsberg1963}
\begin{equation}
\Gamma_{e-ph}=\pi\hbar\lambda\omega_0\qquad.
\end{equation}
Within the energy window between $\hbar\omega_{max}$ and the cross\-over regime, where $\Gamma_{e-e}$ becomes dominant (shaded area in fig.~\ref{fig:2}(a)), the rate of energy dissipation of an electron due to the emission of a phonon with energy $\hbar\omega_0$ is given by 
\begin{equation}
\label{eqn:e-ph}
\frac{\mathrm{d} E}{\mathrm{d} t} = \frac{\hbar\omega_0}{\tau}=\pi\hbar\lambda\omega_0^2\qquad,
\end{equation}
which leads to a linear relaxation of the electron energy~\cite{Gusev1998}. 

The rate of energy relaxation of excited electrons can be extracted from the experimental trARPES intensity $I(\epsilon,t)$ by analyzing the mean excess energy, 
\begin{equation}
\left\langle\Delta E(t)\right\rangle = \frac{\int^{\epsilon_1}_{\epsilon_0}{\epsilon \Delta I(\epsilon,t)\textrm{d}\epsilon}}{\int^{\epsilon_1}_{\epsilon_0}{\Delta I(\epsilon,t)\textrm{d}\epsilon}}\qquad.
\end{equation}
However, this integral represents the mean excess energy within a selected partition of a transient distribution function and not the energy relaxation of single individual electrons. Thus, special care has to be taken with the determination of the integration boundaries, $\epsilon_0$ and $\epsilon_1$. 
A careful investigation of the influence of the integration boundaries and the excitation fluence reveals that a reasonable choice of $\epsilon_0={70}\,\mathrm{meV}$ and $\epsilon_1={300}\,\mathrm{meV}$ allows to give a good estimate of the e-ph coupling strength, as detailed in~\cite{Rettig2012b}.

Experimental data of \Eu, \CoBa and \Ba near $k_\mathrm{F}$ and at $T={100}\,\mathrm{K}$ are shown in fig.~\ref{fig:2}(b). To minimize lattice heating and the influence of the hot thermalized electron distribution, low excitation fluences of $F\sim{50}\,\mathrm{\mu Jcm^{-2}}$ have been used. The linear fits to equation \eqref{eqn:e-ph} within the first ${100}\,\mathrm{fs}$ yield values of $\lambda\left\langle\omega^2\right\rangle={65(5)}\,\mathrm{meV^2}$ for \Eu, ${50(3)}\,\mathrm{meV^2}$ for \CoBa and ${34(6)}\,\mathrm{meV^2}$ for \Ba. Here, error bars represent the numerical uncertainties of the fits, while the overall errors have to be considered larger due to the uncertainty of the integration boundaries discussed above and of the elevated lattice temperature. Nevertheless, we find a good agreement with the values obtained in the 3TM simulations and a considerably higher e-ph coupling in \Eu than in \CoBa  and even lower $\lambda\left\langle\omega^2\right\rangle$ in \Ba. 

\subsection{Temperature-dependent hole relaxation rates}
\begin{figure}[tb]
\begin{center}
\resizebox{0.6\columnwidth}{!}{\includegraphics{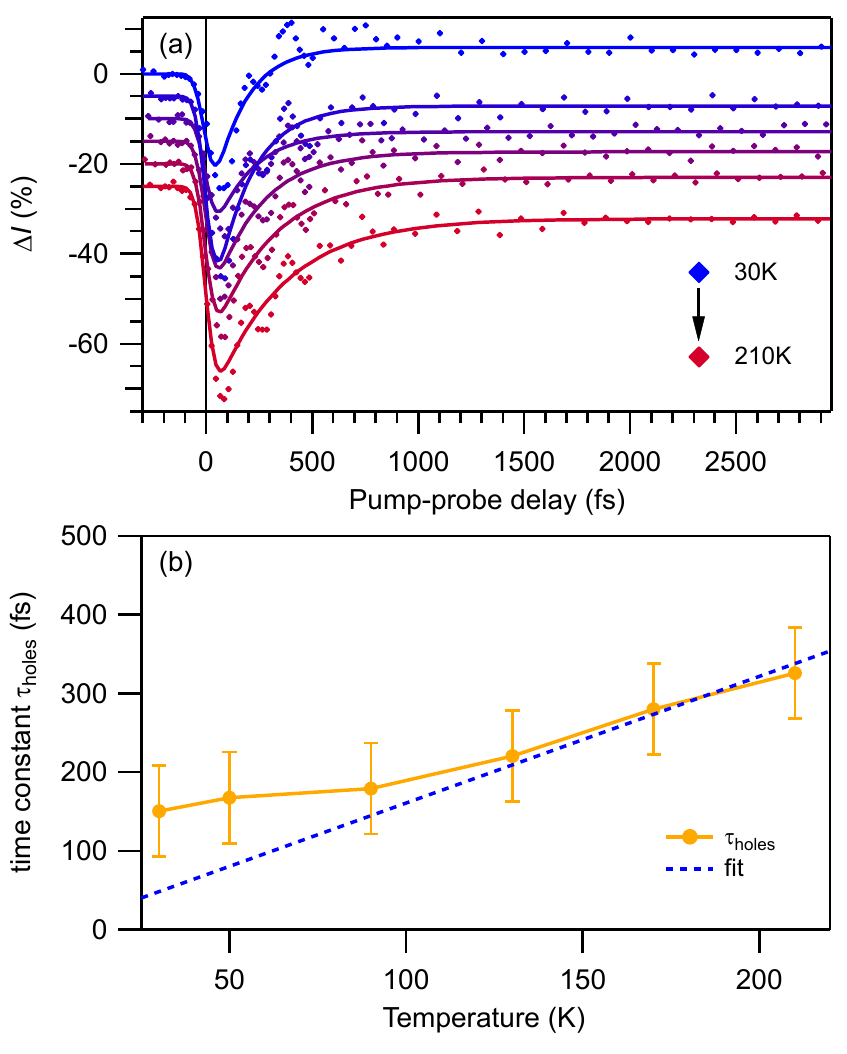}}
\end{center}

\caption{Evaluation of the temperature dependence of relaxation rates in \Eu. (a) Time-dependent spectral weight of holes at  $k_\parallel>k_\mathrm{F}$ for various temperatures. Data are vertically offset for clarity. Solid lines are exponential fits (see text). (b) Relaxation time constants $\tau_\mathrm{holes}$ determined from the data in (a) as function of temperature. The dashed line is a fit to equation~\eqref{eqn:tau_vs_T} in the temperature range $T>100\mathrm{K}$.}
\label{fig:3}   
\end{figure}

Finally, another estimate of the e-ph coupling strength can be gained from the temperature dependence of quasiparticle (QP) relaxation rates~\cite{Allen1987,Gadermaier2010,Kabanov2008,Stojchevska2010}. A recent theoretical investigation of the Boltzmann equation for e-e and e-ph interaction found an analytic solution for the temperature dependence of the QP relaxation rate $\tau$~\cite{Kabanov2008}. In the limit of a bad metal at elevated temperatures, $\tau$ depends linearly on the lattice temperature $T$~\cite{Gadermaier2010,Kabanov2008,Stojchevska2010}:
\begin{equation}
\label{eqn:tau_vs_T}
\tau=\frac{2\pi k_B T}{3\hbar\lambda\left\langle\omega^2\right\rangle}\qquad.
\end{equation}
This relation allows for an independent determination of the e-ph coupling strength $\lambda\left\langle\omega^2\right\rangle$. 

Experimentally, the temperature-dependent relaxation of holes in the hole pocket near $\Gamma$ in \Eu are used, which are found to be independent from the antiferromagnetic transition~\cite{Rettig2012}. The respective hole relaxation time $\tau_\mathrm{holes}$ is determined from the temperature-dependent trARPES intensity by fitting exponential decay curves (fig.~\ref{fig:3}(a)) and is shown in fig.~\ref{fig:3}(b) as a function of temperature. The fit to equation~\eqref{eqn:tau_vs_T} for $T>{100}\,\mathrm{K}$ reveals a good agreement to the linear behavior and yields a value of  $\lambda\left\langle\omega^2\right\rangle={90(30)}\,\mathrm{{meV}^2}$, where the error represents a confidence interval of 80\%. This value, albeit considerably higher than the values found in the 3TM and in the evaluation of the excess energy, is in agreement with the other evaluations within error bars. 

\section{Discussion}
\label{sec:discussion}

The values of $\lambda\left\langle\omega^2\right\rangle$ determined by the three methods discussed above are compared in table~\ref{tab:2}. Despite the limitations and drawbacks of the various methods, we find a perfect agreement of all three methods within error bars. Especially, the trend for larger e-ph coupling in \Eu than in \CoBa and in \Ba is nicely represented. Remarkably, all methods produce a  small $\lambda\left\langle\omega^2\right\rangle<100\,\mathrm{meV^2}$, which already indicates weak e-ph coupling.

Our values of $\lambda\left\langle\omega^2\right\rangle$ can be compared to recently published values determined from optical pump-probe experiments. Mansart \textit{et al.}~\cite{Mansart2010} report a comparable value of $\lambda\left\langle\omega^2\right\rangle\approx{64}\,\mathrm{meV^2}$ for Co-doped \Ba, which was determined using the 3TM. Stojchevska \textit{et al.}~\cite{Stojchevska2010} derived a somewhat higher value of $\lambda\left\langle\omega^2\right\rangle={110(10)}\,\mathrm{meV^2}$ in SrFe$_2$As$_2$ using temperature-dependent QP relaxation times, and for SmFeAsO$_{1-x}$F$_x$ an even larger value of $\lambda\left\langle\omega^2\right\rangle={135(10)}\,\mathrm{meV^2}$ is reported~\cite{Mertelj2010}.

\begin{table}

\caption{Values of $\lambda\left\langle\omega^2\right\rangle$ determined by the three methods.}
\label{tab:2} 
\begin{center}  
\begin{tabular}{l|l|l|l}
	\hline\noalign{\smallskip}
	compound & $T_e(t)$ & $\left\langle\Delta E(t)\right\rangle$ & $\tau_\mathrm{holes}(t)$\\ 
	\noalign{\smallskip}\hline\noalign{\smallskip}
	\Eu & $56-65$ & $65(5)$ & $90(30)$ \\
	\CoBa & $46-55$ & $50(3)$ & - \\
	\Ba & $30-46$ & $34(6)$ & - \\
	\noalign{\smallskip}\hline
	\end{tabular}
\end{center}
\end{table}

Based on our results of $\lambda\left\langle\omega^2\right\rangle$ we can estimate the value of the e-ph coupling constant $\lambda$ for a particular value of $\omega$. Considering the Raman active A$_{1g}$ mode at ${23}\,\mathrm{meV}$, which can be coherently excited~\cite{Avigo2013,Mansart2009,Kim2012} and therefore shows enhanced e-ph coupling in the system, we find $\lambda<0.2$ for all compounds. This estimate is in agreement with calculations of various Fe-pnictide compounds, which report values of $\lambda<0.35$~\cite{Boeri2008,Boeri2010}. Taking the mean of the phonon spectrum as reference, $\lambda$ gets even smaller, in agreement with other publications~\cite{Mansart2010,Stojchevska2010}. Even if we consider the lowest coupled modes around ${12}\,\mathrm{meV}$ to be most important for e-ph coupling, $\lambda$ does not exceed a value of 0.5. Similarly small values for $\lambda$ have been found in the cuprate HTSCs~\cite{Perfetti2007,Gadermaier2010}, which suggests limited importance of e-ph coupling for the pairing mechanism in both classes of materials.

These boundaries for the e-ph coupling constant $\lambda$ allow to estimate the superconducting critical temperature $T_c$, assuming a conventional BCS-type pairing based on e-ph coupling. Here, $T_c$ in isotropic systems and in a strong coupling regime is given by McMillan's formula~\cite{McMillan1968}, modified by Allen and Dynes~\cite{Allen1975},
\begin{equation}
T_c=\frac{\omega_\mathrm{log}}{1.20}\exp\left(-\frac{1.04(1+\lambda)}{\lambda-\mu^*-0.62\lambda\mu^*}\right)
\end{equation}
where $\mu^*$ is the effective Coulomb repulsion and $\omega_\mathrm{log}$ is the logarithmic average of the phonon spectrum. Taking $\mu^*=0$ and $\omega_\mathrm{log}={205}\,\mathrm{K}$~\cite{Boeri2008}, which provides an upper limit for $T_c$, we find $T_c={0.33}\mathrm{K}$ and $T_c={7.5}\mathrm{K}$ for $\lambda=0.2$ and $\lambda=0.5$, respectively. Comparing these results with the experimentally found $T_c=24\,\mathrm{K}$ in optimally doped \CoBa~\cite{Chu2009} and even up to $T_c=38\,\mathrm{K}$ in K doped \Ba~\cite{Rotter2008} demonstrates that e-ph coupling in a conventional BCS pairing scenario cannot explain the high critical temperatures found in the pnictide HTSCs. In addition, albeit the stronger e-ph coupling, the critical temperatures found in \Eu upon Co~\cite{Jiang2009} and K~\cite{Anupam2011} substitution are considerably smaller compared to \Ba.

However, e-ph coupling might still play a significant role in the Cooper pair formation in the Fe pnictides. For instance, a very strong sensitivity of the Fe magnetic moment on the As height in the FeAs tetrahedra was found in DFT band structure calculations~\cite{Yin2008}, with a rate of $6.8\,\mathrm{\mu_B/\textrm{\AA}}$. Inelastic x-ray scattering experiments on CaFe$_2$As$_2$ concluded on strong coupling of phonons to magnetic excitations even in the high-temperature paramagnetic phase~\cite{Hahn2009}. Furthermore, a strong sensitivity of the maximum critical temperature on the pnictogen height was found~\cite{Johnston2010}. On the basis of these findings a strong magneto-phonon coupling in these compounds was proposed and, e.g. the strongly coupled A$_{1g}$ mode perpendicular to the Fe layers, which modulates the pnictogen height, could mediate superconductivity in the Fe-pnictides~\cite{Egami2010}. Such a scenario is supported by the strong coupling of this particular mode to electronic states directly at the Fermi level~\cite{Avigo2013} evidenced by the coherent excitation of this phonon mode~\cite{Avigo2013,Mansart2009,Kim2012}.

\section{Summary}
\label{sec:summary}
In summary, we demonstrated the quantitative an\-a\-ly\-sis of the e-ph coupling strength from femtosecond time- and angle-resolved photoemission spectroscopy. In detail, we compared three different methods to determine the e-ph coupling strength in three 122 FeAs compounds from time- and angle-resolved photoemission experiments. The transient temperature of the thermalized electronic distribution is analyzed in \Eu, \CoBa and \Ba by a three-temperatu\-re model, while the rate of energy relaxation of the non-thermal electrons yields direct information on the second moment of the Eliashberg coupling function $\lambda\left\langle\omega^2\right\rangle$. Finally, the temperature-dependence of hole relaxation rates in \Eu also allows to determine $\lambda\left\langle\omega^2\right\rangle$. All three methods consistently yield a small  $\lambda\left\langle\omega^2\right\rangle<100\,\mathrm{meV^2}$, which results in $\lambda<0.2$ using a reasonable choice of the phonon spectrum. This value is discussed to be too small to explain superconductivity in the Fe pnictides by a conventional BCS-type pairing.

\section{Acknowledgements}
Experiments were carried out at the Freie Universit\"at Berlin. We acknowledge discussions with Martin Wolf and Hermann D\"urr in the early stage of the project and experimental support by Setti Thirupathaiah. This work was supported by the Mercator Research Center Ruhr and the Deutsche Forschungsgemeinschaft through BO 1823/2 and SPP 1458. R.C. acknowledges the Alexander von Humboldt Foundation.

\clearpage
\section{References}
\bibliographystyle{unsrt}
\bibliography{Fe_Pnictides}   

\begin{thebibliography}{10}

\bibitem{Johnston2010}
D.~C. Johnston.
\newblock {T}he puzzle of high temperature superconductivity in layered iron
  pnictides and chalcogenides.
\newblock {\em Adv. Phys.}, 59:803, 2010.

\bibitem{Stewart2011}
G.~R. Stewart.
\newblock {S}uperconductivity in iron compounds.
\newblock {\em Rev. Mod. Phys.}, 83:1589, 2011.

\bibitem{Mazin2008}
I.~I. Mazin, D.~J. Singh, M.~D. Johannes, and M.~H. Du.
\newblock {U}nconventional {S}uperconductivity with a {S}ign {R}eversal in the
  {O}rder {P}arameter of {L}a{F}e{A}s{O}$_{1-x}${F}$_{x}$.
\newblock {\em Phys. Rev. Lett.}, 101:057003, 2008.

\bibitem{Wray2008}
L.~Wray, D.~Qian, D.~Hsieh, et~al.
\newblock {M}omentum dependence of superconducting gap, strong-coupling
  dispersion kink, and tightly bound {C}ooper pairs in the high-${T}_{c}$
  {S}r{B}a$_{1-x}$({K},{N}a)$_{x}${F}e$_{2}${A}s$_{2}$ superconductors.
\newblock {\em Phys. Rev. B}, 78:184508, 2008.

\bibitem{Richard2009}
P.~Richard, T.~Sato, K.~Nakayama, S.~Souma, T.~Takahashi, Y.-M. Xu, G.~F. Chen,
  J.~L. Luo, N.~L. Wang, and H.~Ding.
\newblock {A}ngle-{R}esolved {P}hotoemission {S}pectroscopy of the {F}e-{B}ased
  {B}a$_{0.6}${K}$_{0.4}${F}e$_{2}${A}s$_{2}$ {H}igh {T}emperature
  {S}uperconductor: {E}vidence for an {O}rbital {S}elective {E}lectron-{M}ode
  {C}oupling.
\newblock {\em Phys. Rev. Lett.}, 102:047003, 2009.

\bibitem{Koitzsch2009}
A.~Koitzsch, D.~S. Inosov, D.~V. Evtushinsky, V.~B. Zabolotnyy, A.~A. Kordyuk,
  A.~Kondrat, C.~Hess, M.~Knupfer, B.~B\"uchner, G.~L. Sun, V.~Hinkov, C.~T.
  Lin, A.~Varykhalov, and S.~V. Borisenko.
\newblock {T}emperature and {D}oping-{D}ependent {R}enormalization {E}ffects of
  the {L}ow {E}nergy {E}lectronic {S}tructure of
  {B}a$_{1-x}${K}$_{x}${F}e$_{2}${A}s$_{2}$ {S}ingle {C}rystals.
\newblock {\em Phys. Rev. Lett.}, 102:167001, 2009.

\bibitem{Allen1987}
P.~B. Allen.
\newblock {T}heory of thermal relaxation of electrons in {M}etals.
\newblock {\em Phys. Rev. Lett.}, 59:1460, 1987.

\bibitem{DelFatti2000}
N.~Del~Fatti, C.~Voisin, M.~Achermann, S.~Tzortzakis, D.~Christofilos, and
  F.~Vall\'ee.
\newblock {N}onequilibrium electron dynamics in noble metals.
\newblock {\em Phys. Rev. B}, 61:16956, 2000.

\bibitem{Lisowski2004}
M.~Lisowski, P.~A. Loukakos, U.~Bovensiepen, J.~St\"ahler, C.~Gahl, and
  M.~Wolf.
\newblock {U}ltra-fast dynamics of electron thermalization, cooling and
  transport effects in $\text{Ru(001)}$.
\newblock {\em Appl. Phys. A}, 78:165, 2004.

\bibitem{Bovensiepen2010}
U.~Bovensiepen, H.~Petek, and M.~Wolf, editors.
\newblock {\em {D}ynamics at {S}olid {S}tate {S}urfaces and {I}nterfaces,
  {V}ol. 1}.
\newblock Wiley-VCH Weinheim, Germany, 2010.

\bibitem{Bovensiepen2012}
U.~Bovensiepen and P.~S. Kirchmann.
\newblock {E}lementary relaxation processes investigated by femtosecond
  photoelectron spectroscopy of two-dimensional materials.
\newblock {\em Laser \& Photonics Reviews}, 6(5):589, 2012.

\bibitem{Anisimov1974}
S.~I. Anisimov, B.~L. Kapeliovich, and T.~L. Perelman.
\newblock {E}lectron {E}mission from {M}etal {S}urfaces {E}xposed to
  {U}ltrashort {L}aser {P}ulses.
\newblock {\em Sov. Phys. JETP-USSR}, 39:375, 1974.

\bibitem{Rethfeld2002}
B.~Rethfeld, A.~Kaiser, M.~Vicanek, and G.~Simon.
\newblock {U}ltrafast {D}ynamics of {N}onequilibrium {E}lectrons in {M}etals
  {U}nder {F}emtosecond {L}aser {I}rradiation.
\newblock {\em Phys. Rev. B}, 65:214303, 2002.

\bibitem{Perfetti2007}
L.~Perfetti, P.~A. Loukakos, M.~Lisowski, U.~Bovensiepen, H.~Eisaki, and
  M.~Wolf.
\newblock {U}ltrafast electron relaxation in superconducting
  {B}i$_2${S}r$_2${C}a{C}u$_2${O}$_{8 + \delta}$ by time-resolved photoelectron
  spectroscopy.
\newblock {\em Phys. Rev. Lett.}, 99:197001, 2007.

\bibitem{Gadermaier2010}
C.~Gadermaier, A.~S. Alexandrov, V.~V. Kabanov, P.~Kusar, T.~Mertelj, X.~Yao,
  C.~Manzoni, D.~Brida, G.~Cerullo, and D.~Mihailovic.
\newblock {E}lectron-{P}honon {C}oupling in {H}igh-{T}emperature {C}uprate
  {S}uperconductors {D}etermined from {E}lectron {R}elaxation {R}ates.
\newblock {\em Phys. Rev. Lett.}, 105:257001, 2010.

\bibitem{Mansart2010}
B.~Mansart, D.~Boschetto, A.~Savoia, F.~Rullier-Albenque, F.~Bouquet,
  E.~Papalazarou, A.~Forget, D.~Colson, A.~Rousse, and M.~Marsi.
\newblock {U}ltrafast transient response and electron-phonon coupling in the
  iron-pnictide superconductor
  $\text{Ba(Fe}_{1-x}\text{Co}_x\text{)}_2\text{As}_2$.
\newblock {\em Phys. Rev. B}, 82:024513, 2010.

\bibitem{Avigo2013}
I.~Avigo, R.~Cortés, L.~Rettig, S.~Thirupathaiah, H.~S. Jeevan, P.~Gegenwart,
  T.~Wolf, M.~Ligges, M.~Wolf, J.~Fink, and U.~Bovensiepen.
\newblock {C}oherent excitations and electron-phonon coupling in
  {B}a/{E}u{F}e$_2${A}s$_2$ compounds investigated by femtosecond time- and
  angle-resolved photoemission spectroscopy.
\newblock {\em J. Phys.: Condens. Matter}, 25(9):094003, 2013.

\bibitem{Echenique2000}
P.~M. Echenique, J.~M. Pitarke, E.~V. Chulkov, and A.~Rubio.
\newblock {T}heory of {I}nelastic {L}ifetimes of {L}ow-{E}nergy {E}lectrons in
  {M}etals.
\newblock {\em Chem. Phys.}, 251:1, 2000.

\bibitem{McDougall1995}
B.~A. McDougall, T.~Balasubramanian, and E.~Jensen.
\newblock {P}honon contribution to quasiparticle lifetimes in {C}u measured by
  angle-resolved photoemission.
\newblock {\em Phy. Rev. B}, 51:13891, 1995.

\bibitem{Valla1999}
T.~Valla, A.~V. Fedorov, P.~D. Johnson, and S.~L. Hulbert.
\newblock {M}any-body effects in angle-resolved photoemission: {Q}uasiparticle
  energy and lifetime of a {M}o(110) surface state.
\newblock {\em Phys. Rev. Lett.}, 83:2085, 1999.

\bibitem{Engelsberg1963}
S.~Engelsberg and J.~R. Schrieffer.
\newblock {C}oupled {E}lectron-{P}honon {S}ystem.
\newblock {\em Phys. Rev.}, 131:993, 1963.

\bibitem{Gusev1998}
V.~E. Gusev and O.~B. Wright.
\newblock {U}ltrafast nonequilibrium dynamics of electrons in metals.
\newblock {\em Phys. Rev. B}, 57:2878, 1998.

\bibitem{Rettig2012b}
L.~Rettig.
\newblock {\em {U}ltrafast {D}ynamics of {C}orrelated {E}lectrons}.
\newblock PhD thesis, Freie Universit\"at Berlin, 2012.

\bibitem{Kabanov2008}
V.~V. Kabanov and A.~S. Alexandrov.
\newblock {E}lectron relaxation in metals: {T}heory and exact analytical
  solutions.
\newblock {\em Phys. Rev. B}, 78:174514, 2008.

\bibitem{Stojchevska2010}
L.~Stojchevska, P.~Kusar, T.~Mertelj, V.~V. Kabanov, X.~Lin, G.~H. Cao, Z.~A.
  Xu, and D.~Mihailovic.
\newblock {E}lectron-phonon coupling and the charge gap of spin-density wave
  iron-pnictide materials from quasiparticle relaxation dynamics.
\newblock {\em Phys. Rev. B}, 82:012505, 2010.

\bibitem{Rettig2012}
L.~Rettig, R.~Cort\'es, S.~Thirupathaiah, P.~Gegenwart, H.~S. Jeevan, M.~Wolf,
  J.~Fink, and U.~Bovensiepen.
\newblock {U}ltrafast {M}omentum-{D}ependent {R}esponse of {E}lectrons in
  {A}ntiferromagnetic $\text{EuFe}_2\text{As}_2$ {D}riven by {O}ptical
  {E}xcitation.
\newblock {\em Phys. Rev. Lett.}, 108:097002, 2012.

\bibitem{Mertelj2010}
T.~Mertelj, P.~Kusar, V.~V. Kabanov, L.~Stojchevska, N.~D. Zhigadlo,
  S.~Katrych, Z.~Bukowski, J.~Karpinski, S.~Weyeneth, and D.~Mihailovic.
\newblock {Q}uasiparticle relaxation dynamics in spin-density-wave and
  superconducting $\text{SmFeAsO}_{1-x}\text{F}_x$ single crystals.
\newblock {\em Phys. Rev. B}, 81:224504, 2010.

\bibitem{Mansart2009}
B.~Mansart, D.~Boschetto, A.~Savoia, F.~Rullier-Albenque, A.~Forget, D.~Colson,
  A.~Rousse, and M.~Marsi.
\newblock {O}bservation of a coherent optical phonon in the iron pnictide
  superconductor $\text{Ba}\text{(Fe}_{1 - x}\text{Co}_x\text{)}_2\text{As}_2$
  ($x$ = 0.06 and 0.08).
\newblock {\em Phys. Rev. B}, 80:172504, 2009.

\bibitem{Kim2012}
K.~W. Kim, A.~Pashkin, H.~Sch\"afer, M.~Beyer, M.~Porer, T.~Wolf, C.~Bernhard,
  J.~Demsar, R.~Huber, and A.~Leitenstorfer.
\newblock {U}ltrafast transient generation of spin-density-wave order in the
  normal state of $\text{BaFe}_2\text{As}_2$ driven by coherent lattice
  vibrations.
\newblock {\em Nat. Mater.}, 11:497, 2012.

\bibitem{Boeri2008}
L.~Boeri, O.~V. Dolgov, and A.~A. Golubov.
\newblock {I}s $\text{LaFeAsO}_{1-x}\text{F}_x$ an {E}lectron-{P}honon
  {S}uperconductor?
\newblock {\em Phys. Rev. Lett.}, 101:026403, 2008.

\bibitem{Boeri2010}
L.~Boeri, M.~Calandra, I.~I. Mazin, O.~V. Dolgov, and F.~Mauri.
\newblock {E}ffects of magnetism and doping on the electron-phonon coupling in
  $\text{BaFe}_{2}\text{As}_{2}$.
\newblock {\em Phys. Rev. B}, 82:020506, 2010.

\bibitem{McMillan1968}
W.~L. McMillan.
\newblock {T}ransition {T}emperature of {S}trong-{C}oupled {S}uperconductors.
\newblock {\em Phys. Rev.}, 167:331, 1968.

\bibitem{Allen1975}
P.~B. Allen and R.~C. Dynes.
\newblock {T}ransition temperature of strong-coupled superconductors
  reanalyzed.
\newblock {\em Phys. Rev. B}, 12:905, 1975.

\bibitem{Chu2009}
J.-H. Chu, J.~G. Analytis, C.~Kucharczyk, and I.~R. Fisher.
\newblock {D}etermination of the phase diagram of the electron-doped
  superconductor {B}a({F}e$_{1-x}${C}o$_x$)$_{2}${A}s$_{2}$.
\newblock {\em Phy. Rev. B}, 79:014506, 2009.

\bibitem{Rotter2008}
M.~Rotter, M.~Tegel, D.~Johrendt, I.~Schellenberg, W.~Hermes, and R.~P\"ottgen.
\newblock {S}pin-density-wave anomaly at 140 {K} in the ternary iron arsenide
  ${\text{bafe}}_{2}{\text{as}}_{2}$.
\newblock {\em Phys. Rev. B}, 78:020503, 2008.

\bibitem{Jiang2009}
S.~Jiang, H.~Xing, G.~Xuan, Z.~Ren, C.~Wang, Z.-a. Xu, and G.~Cao.
\newblock {S}uperconductivity and local-moment magnetism in
  $\text{Eu}{({\text{Fe}}_{0.89}{\text{Co}}_{0.11})}_{2}{\text{as}}_{2}$.
\newblock {\em Phys. Rev. B}, 80:184514, 2009.

\bibitem{Anupam2011}
Anupam, P.~L. Paulose, S.~Ramakrishnan, and Z.~Hossain.
\newblock {D}oping dependent evolution of magnetism and superconductivity in
  {E}u$_{1-x}${K}$_x${F}e$_2${A}s$_2$ (x=0-1) and temperature dependence of the
  lower critical field {H}$_{c1}$.
\newblock {\em J. Phys.: Condens. Matter}, 23:455702, 2011.

\bibitem{Yin2008}
Z.~P. Yin, S.~Leb\`egue, M.~J. Han, B.~P. Neal, S.~Y. Savrasov, and W.~E.
  Pickett.
\newblock {E}lectron-{H}ole {S}ymmetry and {M}agnetic {C}oupling in
  {A}ntiferromagnetic {L}a{F}e{A}s{O}.
\newblock {\em Phys. Rev. Lett.}, 101:047001, 2008.

\bibitem{Hahn2009}
S.~E. Hahn, Y.~Lee, N.~Ni, P.~C. Canfield, A.~I. Goldman, R.~J. McQueeney,
  B.~N. Harmon, A.~Alatas, B.~M. Leu, E.~E. Alp, D.~Y. Chung, I.~S. Todorov,
  and M.~G. Kanatzidis.
\newblock {I}nfluence of magnetism on phonons in $\text{CaFe}_2 \text{As}_2$ as
  seen via inelastic x-ray scattering.
\newblock {\em Phys. Rev. B}, 79:220511, 2009.

\bibitem{Egami2010}
T.~Egami, B.~V. Fine, D.~Parshall, A.~Subedi, and D.~J. Singh.
\newblock {S}pin-{L}attice {C}oupling and {S}uperconductivity in {F}e
  {P}nictides.
\newblock {\em Adv. Cond. Matt. Phys.}, 2010:164916, 2010.

\end{thebibliography}

\end{document}